\documentclass[11pt,twoside]{article}
\usepackage{CAGN2021}
\usepackage{graphicx}

\usepackage[T1]{fontenc} 

\usepackage{latexsym}
\usepackage{verbatim}

\usepackage{ifpdf}  
\ifpdf  
      \DeclareGraphicsExtensions{.pdf,.png,.jpg}  
\else  
      \DeclareGraphicsExtensions{.eps}  
\fi 

\usepackage{xspace}
\newcommand{\hii}{H~{\scshape ii}\xspace}
\newcommand{\Ha}{\ifmmode {\mathrm{H}\alpha} \else H$\alpha$\fi\xspace}
\newcommand{\Hb}{\ifmmode {\mathrm{H}\beta} \else H$\beta$\fi\xspace}
\newcommand{\oii}{\ifmmode [\text{O}\,\textsc{ii}] \else [O~{\scshape ii}]\fi\xspace}
\newcommand{\oiii}{\ifmmode [\text{O}\,\textsc{iii}] \else [O~{\scshape iii}]\fi\xspace}
\newcommand{\Oiii}{\ifmmode [\text{O}\,\textsc{iii}]\lambda 5007 \else [O~{\scshape iii}]$\lambda 5007$\fi\xspace}
\newcommand{\nii}{\ifmmode [\text{N}\,\textsc{ii}] \else [N~{\scshape ii}]\fi\xspace}
\newcommand{\Nii}{\ifmmode [\text{N}\,\textsc{ii}]\lambda 6584 \else [N~{\scshape ii}]$\lambda 6584$\fi\xspace}
\newcommand{\sii}{\ifmmode [\text{S}\,\textsc{ii}] \else [S~{\scshape ii}]\fi\xspace}
\newcommand{\Sii}{\ifmmode [\text{S}\,\textsc{ii}]\lambda 6716 \else [S~{\scshape ii}]$\lambda 6716$\fi\xspace}
\newcommand{\WHa}{\ifmmode W_{\mathrm{H}\alpha} \else $W_{\mathrm{H}\alpha}$\fi\xspace}

\begin{document}

\vskip 1.0cm
\markboth{N.~Vale Asari}{The role of the DIG in metallicity calibrations}
\pagestyle{myheadings}
%
%
\vspace*{0.5cm}
\parindent 0pt{Contributed Paper}


\vspace*{0.5cm}
\title{The role of the diffuse ionized gas in metallicity calibrations}

\author{N.~Vale Asari$^{1, 2}$}
\affil{$^1$Depto.\ de F\'isica -- CFM -- UFSC, Florian\'opolis, SC, Brazil \\ 
  $^2$Royal Society--Newton Advanced Fellowship}

\begin{abstract}
  Estimates of gas-phase abundances based on strong-line methods have
  been calibrated for \hii regions.  Those methods ignore any
  contribution from the diffuse ionized gas (DIG), which shows
  enhanced collisional-to-recombination line ratios in comparison to
  \hii regions of the same metallicity. Applying strong line methods
  whilst ignoring the role of the DIG thus systematically
  overestimates metallicities. Using integral field spectroscopy data,
  we show how to correct for the DIG contribution and how it biases
  the mass--metallicity--star formation rate relation.


\end{abstract}

\section{Introduction}


The diffuse ionized gas (DIG) is a warm ($10^4$ K), low-density
($10^{-1}$ cm$^{-3}$) gas phase in the interstellar medium. The DIG
was first detected outwith the plane of the Milky Way, and later on as
extraplanar emission in other galaxies, in interarm regions, and in
bulges of galaxies \citep[e.g.][]{Hoyle.Ellis.1963a, Dettmar.1990a,
  Gomes.etal.2016c}.  Apart from their lower densities compared to
\hii regions, the DIG has been found to have higher electronic
temperatures and enhanced collisional-to-recombination line ratios,
such as \Nii/\Ha, \Sii/\Ha, and for some objects \Oiii/\Hb.  For a
thorough review of the DIG, see \citet{Haffner.etal.2009a} and
references therein.


The enhanced collisional-to-recombination line ratios imply that the
DIG must be ionized by a source that is harder than OB stars. Several
sources have been proposed, such as shocks from supernova winds,
turbulent dissipation, photoelectric heating by grains, leakage of
photons from \hii regions, and hot low mass evolved stars (HOLMES,
\citealp{Stasinska.etal.2008a, FloresFajardo.etal.2011a}).

Regardless of its ionization mechanism, the presence of the DIG biases
commonly used estimates of gas-phase abundances.  Consider a
hypothetical galaxy with a single value of oxygen abundance (O/H).
The presence of the DIG would enhance \nii/\Ha.  If we were to obtain
O/H from the \nii/\Ha index using a strong line method, which has been
calibrated for \hii regions, we would wrongly conclude this galaxy has
a higher O/H than it really does.  The larger the contribution of the
DIG, the more O/H is overestimated when using strong-line methods.  We
investigate the effect of the DIG on integrated values of O/H based on
integral field spectroscopy (IFS) from Mapping Nearby Galaxies at APO
(MaNGA, \citealp{Blanton.etal.2017a}). \citet{ValeAsari.etal.2019a},
hereafter VA19, detail the sample selection and data processing. The
next sections summarize their main findings.

\section{Identifying the DIG}

\begin{figure}[tb]
  \begin{center}
    \includegraphics[width=0.59\textwidth, trim=0 0 0 50]{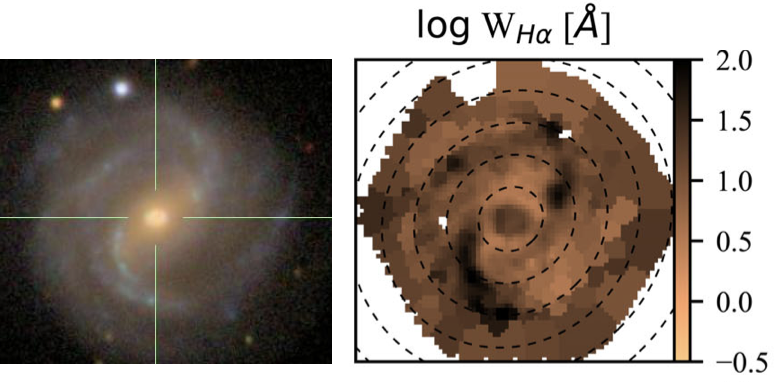} 
    \caption{ Right: optical SDSS image of galaxy NGC 0776. Left: map
      of \Ha equivalent width of the same galaxy based on CALIFA
      observations. High values of $W_\Ha$ trace the spiral arms,
      denoting its usefulness in separating emission dominated by \hii
      regions from emission dominated by the DIG in the bulge and
      interarm regions.  (Figure adapted from figure~1 by
      \citealp{Lacerda.etal.2018a}.)  }
    \label{WHa-map-L18}
  \end{center}
\end{figure}

\begin{figure}[tb]
  \begin{center}
    \includegraphics[width=0.39\textwidth]{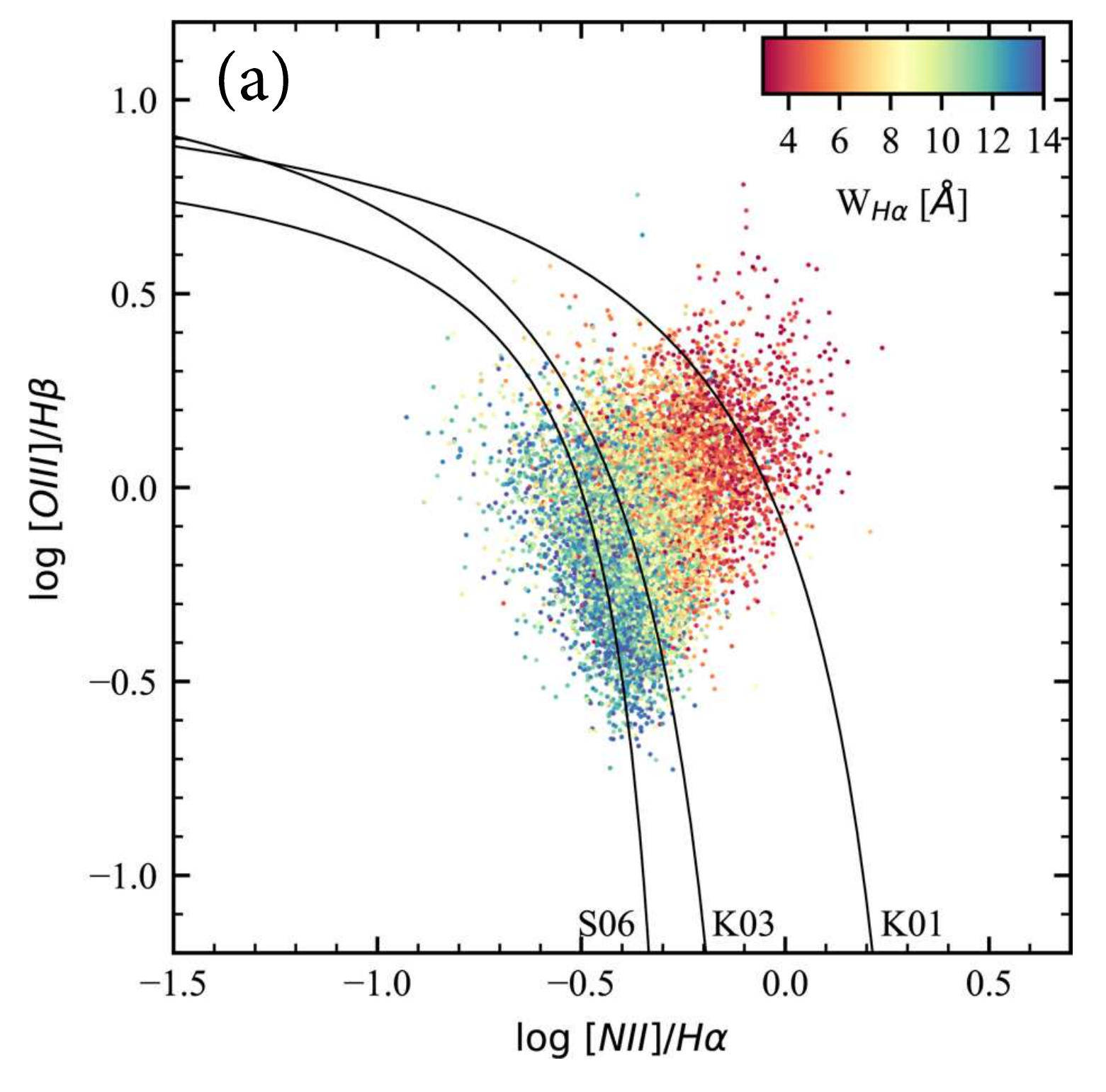} 
    \includegraphics[width=0.6\textwidth]{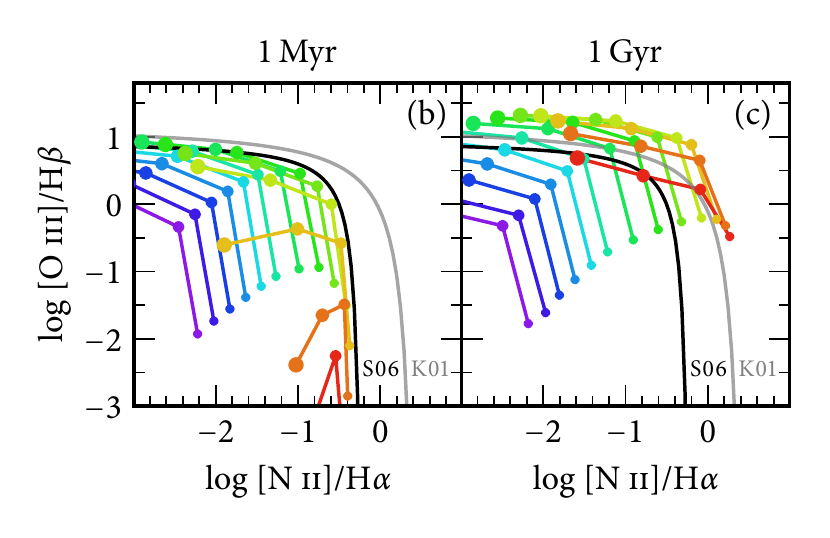} 
    \caption{ \textbf{(a)} Spaxels classified as mDIG from 391 CALIFA
      galaxies on the \nii/\Ha versus \oiii/\Hb plane, colour-coded by
      $W_\Ha$. Lines marked as S06, K03 and K01 are the curves by
      \citet{Stasinska.etal.2006a, Kauffmann.etal.2003c,
        Kewley.etal.2001a}.  S06 is a line based on photoionization
      models calibrated on SDSS to the right of which contribution
      from radiation harder than OB stars is needed to explain the
      observed emission-line ratios.  Regions where $W_\Ha$ is smaller
      (red points), thus where the contribution of the DIG is larger,
      have enhanced emission-line ratios. This implies that, the
      larger the contribution from the DIG, the larger the systematic
      changes in emission line ratios.  \textbf{(b), (c)}
      Photoionization models for
      $12 + \log \mathrm{O/H} = 7.0, 7.2, \dots, 9.4$ (from violet to
      red) and $\log U = -4, -3, -2, -1$ (from small to big points)
      for (b) a young SSP, representative of \hii regions, and (c) an
      old SSP, mimicking ionization by HOLMES or another hard
      source. For the same value of O/H and $U$, \nii/\Ha and
      \oiii/\Hb may be enhanced by more than 1.0 dex due to the harder
      ionizing field. (Panel (a) adapted from figure~9 by
      \citealp{Lacerda.etal.2018a}.) }
    \label{BPT-L18}
  \end{center}
\end{figure}

Three main ways have been used to identify regions dominated by the
DIG in IFS data: using a threshold for (1) the \Ha surface brightness,
(2) a collisional-to-recombination line ratio such as \sii/\Ha, and
(3) the \Ha equivalent width ($W_\Ha$). A cut in \Ha surface
brightness \citep[e.g.][]{Zhang.etal.2017c} hinges on the fact that
the DIG is less dense than \hii regions. This criterion is not
appropriate for central parts of galaxies, as argued by
\citet{Lacerda.etal.2018a}, since the projected surface brightness
would be enhanced. $W_\Ha$, being an intensive quantity, remains small
and can correctly identify DIG-dominated regions even in bulges. A cut
in \sii/\Ha \citep[e.g.][]{Kaplan.etal.2016a} would be inappropriate
here, since we aim to quantify the effect of the DIG on the very same
emission lines. We thus find $W_\Ha$ to be the best physically
motivated criterion for our dataset.


Figure~\ref{WHa-map-L18} shows an optical image of NGC 0776 and a map
of its observed $W_\Ha$ obtained from the Calar Alto Legacy Integral Field
Area Survey (CALIFA; \citealp{Sanchez.etal.2016a}).
High-$W_\Ha$ spaxels trace the spiral arms, whereas small-$W_\Ha$ spaxels
are in the interarm regions dominated by the DIG. The reasoning behind using
$W_\Ha$ is that nebulae ionized by \hii regions have large values of
$W_\Ha$, whereas those ionized by HOLMES (see figure 2 by
\citealp{CidFernandes.etal.2011a}) or e.g.\ by hard photons leaking from
\hii regions would have small $W_\Ha$ values.  Implicit in the $W_\Ha$
criterion is that the ionized nebulae and the source of ionization
have been observed in the same spectrum. Therefore, care is needed
when applying it to high-resolution data (e.g. MUSE,
\citealp{Bacon.etal.2010a}, or SITELLE,
\citealp{Brousseau.etal.2014a}), where the ionizing source and the
nebulae may not be cospatial.

For both CALIFA and MaNGA data (\citealp{Lacerda.etal.2018a}; VA19),
we have classified spaxels as (a) hDIG ($W_\Ha < 14$~\AA),
i.e.\ DIG compatible with ionization by HOLMES; (b) mDIG
($3 < W_\Ha < 14$~\AA), for DIG ionized by mixed processes; and (c)
SFc ($W_\Ha > 14$~\AA), which are dominated by emission from
star-forming complexes.

Figure~\ref{BPT-L18} (a) shows spaxels from CALIFA galaxies classified
as mDIG in the \nii/\Ha versus \oiii/\Hb plane. Spaxels colour-coded
by $W_\Ha$ reveal that, the smaller the $W_\Ha$ value, the more
enhanced \nii/\Ha and \oiii/\Hb are. This denotes $W_\Ha$ can be used
as a proxy for how much the DIG affects collisional-to-recombination
emission line ratios.  Panels (b) and (c) of Figure~\ref{BPT-L18} show
a sequence of photoionization models run with \texttt{cloudy} v.~17.01
\citep{Ferland.etal.2017a}. Both panels display the same sequence of
O/H and ionization parameter ($U$); panel (b) shows ionization by an 1
Myr (= OB stars) and panel (c) by an 1 Gyr (= HOLMES; but can viewed as a
proxy for a harder ionizing source) simple stellar population
\citep{Molla.GarciaVargas.Bressan.2009a}. This exemplifies how, given
a nebula with the same O/H and $U$, \nii/\Ha and \oiii/\Hb are
enhanced if the ionizing field is harder than that of OB stars.

\section{Removing the DIG}

\begin{figure}[tb]
\begin{center}
  \includegraphics[width=\textwidth, trim=0 15 0 20]{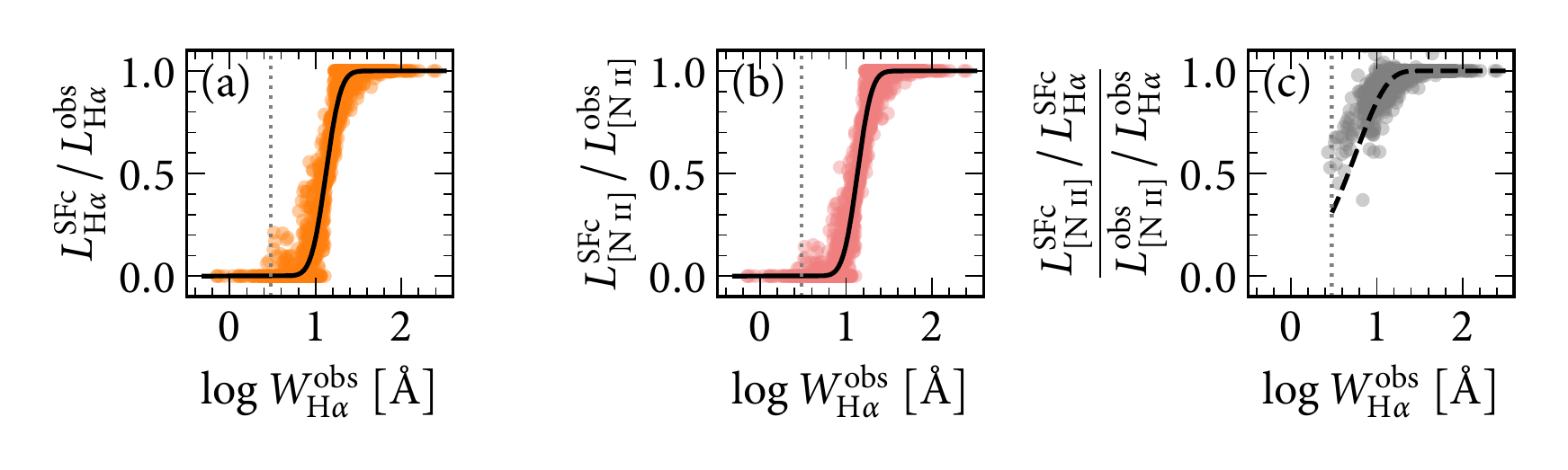}
  \caption{ (a) Fraction of \Ha luminosity in SFc spaxels as a
    function of the global $W_\Ha$ for 1409 MaNGA galaxies in a
    circular aperture corresponding to the $0.7$ effective radius. The
    solid line is a fit to the data points, and is suitable to correct
    fibre-spectroscopy data surveys similar to the SDSS.  (b) The same
    plot for \nii luminosity. The correction for \nii is sharper than
    for \Ha, since the DIG being removed has larger values of
    \nii/\Ha.  (c) Compound effect of the correction shown in (a) and
    (b) to the \nii/\Ha line ratio (dashed line). Grey points show
    \nii/\Ha for SFc spaxels divided by total \nii/\Ha.  Vertical
    dotted lines on all panels mark the $W_\Ha > 3$ \AA\ threshold for
    which correction is valid.  (Panels (a) and (b) adapted from
    figure~3 by VA19.)  }
  \label{DIGcor-V19}
\end{center}
\end{figure}

\begin{figure}[tb]
\begin{center}
  \includegraphics[width=1\textwidth, trim=50 0 220 0, clip]{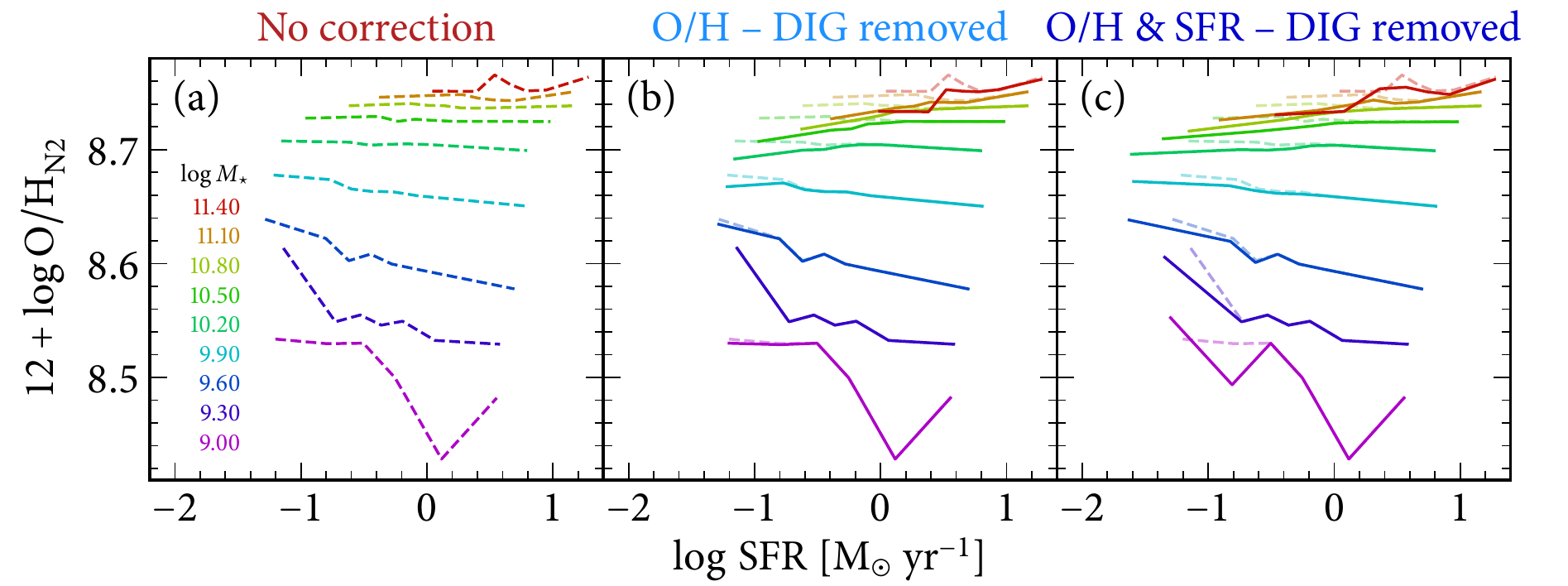}
  \caption{ (a) Stellar mass--metallicity--star formation rate
    relation for $94\,335$ star-forming SDSS galaxies, where O/H has
    been estimated from the \nii/\Ha index. (b) The dashed lines from
    panel (a) are repeated for comparison. The solid lines show
    $M$--O/H--SFR after removing the DIG contribution to \nii and \Ha
    prior to computing O/H using the curves from
    Figure~\ref{DIGcor-V19}. The DIG affects mostly high-mass, low-SFR
    objects; the flat relations for high-mass bins thus become
    positive correlations.  (c) Similar to panel (b), but after removing
    the contribution of the DIG for \Ha and \Hb before computing the
    SFR.
    The effect of the DIG is small because the correction curves from
    Figure~\ref{DIGcor-V19} are based on MaNGA data; at $\sim 1$ kpc
    resolution, MaNGA SFc spaxels are still a mix of DIG and
    \hii-region emission.
    (Figure adapted from figure~5 by VA19.)  }
  \label{MZSFR-V19}
\end{center}
\end{figure}

To correct a given galaxy for the DIG emission, VA19 use the $W_\Ha$
criterion to identify SFc spaxels.  Figure~\ref{DIGcor-V19}(a) shows,
for 1409 star-forming MaNGA galaxies, the quotient of the \Ha
luminosity due only to SFc spaxels to the total \Ha luminosity within
a $0.7$ effective radius. This quotient is plotted as a function of
the global $W_\Ha$ of the galaxy: those galaxies with lower global
$W_\Ha$ have more spaxels tagged as hDIG or mDIG, so a greater
fraction of the \Ha luminosity is removed when removing DIG spaxels.
Figure~\ref{DIGcor-V19}(b) shows the same but for the \nii luminosity.
Crucially, the `DIG-correction' curves for \nii and \Ha are slightly
different: removing the DIG means removing more light from \nii than
from \Ha. Since the curves in panels (a) and (b) seem almost
undistinguishable by eye (but check on table 3 VA19 that they are not
exactly the same), panel (c) shows the compound effect of the DIG
correction on the \nii/\Ha ratio. The $y$-axis is now the \nii/\Ha
ratio summing up only SFc spaxels divided by the total \nii/\Ha ratio.
The dashed line shows the ratio of the curve on panel (b) to the one
on panel (a). One can now clearly see how the DIG correction for
\emph{line luminosities}, although very similar visually, does imply a
change in emission line ratios consistent with
the emission line ratio variation due to DIG for individual MaNGA
galaxies.

Those DIG-correction curves were calculated for a $0.7$ effective
radius, which is the typical coverage of galaxies in the
fibre-spectroscopic data of the Sloan Digital Sky Survey (SDSS;
\citealp{York.etal.2000a}).  We selected $94\,335$ star-forming
galaxies from the SDSS to check the effect of this correction on O/H
and SFR. In the following, we calculate O/H using the \nii/\Ha index
calibrated by \cite{Curti.etal.2017a}. Results for calibrations based
on other emission line indices are shown by VA19.

Figure~\ref{MZSFR-V19} compares the mass--metallicity--star formation
rate ($M$--O/H--SFR) relation without any correction, and after
removing the DIG contribution. Panel (a) shows the relation with no
correction. There is an anti-correlation between O/H and SFR for
low-$M$ bins, and no correlation for high-$M$ bins. This result is
very similar to the original relation found by
\cite{Mannucci.etal.2010a}. Panel (b) repeats the relation in panel
(a) in dashed lines, and overplots the new relation after removing the
contribution from the DIG to O/H. Galaxies with large $M$ and small
SFR are the most affected by the DIG, and so much so that for high-$M$
bins the previously flat correlations become positive correlations.
Panel (c) is similar to panel (b), but we have also removed the
contribution from the DIG to the \Ha and \Hb luminosities prior to the
computation of the SFR.

Those results show that, because strong line methods to measure
abundances are calibrated for \hii regions, it is dangerous to apply
them carelessly when there is an important contribution from the DIG
to a galaxy spectrum. The DIG causes O/H values to be systematically
overestimated.

We warn that the effect shown here and in VA19 is small because SFc
spaxels in MaNGA are not completely devoid of DIG contribution. At the
spatial resolution of $\sim 1$~kpc, they are much larger than
classical \hii regions. A large sample of higher-resolution
spectroscopic data would be needed in order to reveal whether the
correction is larger when considering DIG-free \hii regions.

\acknowledgments
NVA thanks Gra\.zyna Stasi\'nska and
  Ariel Werle for suggestions in the manuscript, and acknowledges
  support of FAPESC, CNPq, and the Royal Society--Newton
  Advanced Fellowship award
  (NAF\textbackslash{}R1\textbackslash{}180403).

\bibliographystyle{aaabib}
\bibliography{references}

\end{document}